# Modulation of Hanle magnetoresistance in an ultrathin platinum film by ionic gating

Yuu Maruyama[1], Ryo Ohshima[1], Ei, Shigematsu[1], Yuichiro Ando,[1] and Masashi Shiraishi[1]*

[1]*Graduate School of Engineering, Department of Electronic Science and Engineering, Kyoto University, Kyoto 615-8510, Japan*

E-mail: shiraishi.masashi.4w@kyoto-u.ac.jp

**Abstract**

Hanle magnetoresistance (HMR) is a type of magnetoresistance where interplay of the spin Hall effect, Hanle-type spin precession, and spin-dependent scattering at the top/bottom surfaces in a heavy metal controls the effect. In this study, we modulate HMR in ultrathin Pt by ionic gating, where the surface Rashba field created by a strong electric field at the interface between the ionic gate and Pt plays the dominant role in the modulation. This finding can facilitate investigations of gate-tunable, spin-related effects and fabrication of spin devices.

Carrier doping is one of the most promising approaches for investigating phenomena and modulating effects in condensed matter physics since the development of field effect transistors consisting of germanium. The key for such efforts is a shift of the Fermi level in a solid that is adjacent to a gate material by gating-attributable carrier accumulation. Although carrier doping by gating is well-established, introducing ionic gating further facilitates condensed matter physics because ionic gating enables quite dense carrier accumulation ($> 10^{14}$ cm$^{-2}$); which facilitates a substantial shift of the Fermi level, resulting in unprecedented physical phenomena. A pioneering work was observation of the insulator–superconductor transition in SrTiO$_3$ by ionic gating [1]; subsequent achievements of the paramagnetic–ferromagnetic transition in a diluted magnetic semiconductor [2], modulation of the Curie temperature in ultrathin Co [3,4], a gate-tunable inverse/ordinal spin Hall effect in ultrathin Pt as well as Pd via tuning of the spin–orbit interaction (SOI) [5-7], and tuning of magnetic proximity [8] substantially expanded the fields of spintronics and spin-orbitronics.

Accordingly, achieving a spintronic effect by using ionic gating has been an active area of research. Given that carrier accumulation and possible creation of the Rashba SOI occur at the interface of a solid and gate insulator, an effect that is sensitive to the surface electronic states is a potential candidate for modulation by gating. Hanle magnetoresistance (HMR) [9-11] is a type of a magnetoresistance effect; where interplay of the spin Hall effect (SHE) [12], the Hanle-type spin precession effect [13], and spin-dependent scattering at surfaces facilitates magnetoresistance in a material with sizable SOI (such as Pt). Spin-dependent scattering at the top and bottom surface of the SOI material determines the manifestation and amplitude of HMR. Hence, modulation of the SOI and an electronic state at a surface by gating-attributable carrier accumulation can tune HMR, which extends the applications of spintronic effects. In this study, we demonstrate gate-tunable HMR in ultrathin Pt by using ionic gating. Ultrathin Pt is an appropriate material for modulating conductivity and SOI because of its considerably low carrier density [5,14,15]. Although this characteristic invokes tunable top surface conductivity in an ionic-gate/ultrathin-Pt bilayer and generation of pure spin current via the SHE, the interplay of these effects contributing to HMR is intricate and inconclusive. Our investigation can facilitate understanding of HMR, one of the pivotal effects that clarifies the coupling between charge and spin currents in a solid with strong SOI.

Figure 1(a) shows the measurement setup of gate-tunable HMR. A 5-μm-thick Y$_3$Fe$_5$O$_{12}$ (yttrium-iron-garnet, YIG) film grown on (111) gadolinium gallium garnet (GGG) (Granopt, Japan) was used as a substrate for growth of ultrathin Pt. The substrate was annealed at 1273 K in air for 90 min for surface cleaning. A 1.7-nm-thick Pt film (ultrathin Pt film) was deposited onto the substrate by sputtering (growth rate: 0.06 nm/s). Electron beam (EB) lithography was used for patterning the Hall-bar [30-μm width (*W*), 70-μm length (*L*)]. Ti(5 nm)/Au(100 nm) electrodes, formed by EB deposition, were equipped to the Hall-bar. An ionic gel was prepared by mixing polystyrene–polymethyl methacrylate–polystyrene (PS-PMMA-PS, Polymer Source, Canada), diethylmethyl(2-methoxyethyl)ammonium bis(trifluoromethylsulfonyl)imide (DEME-TFSI) ionic liquid (Kanto Chemical, Japan), and ethyl propionate ($CH_3CH_2COOC_2H_5$, Nacalai Tesque, Japan) at a mass ratio of 93:7:200; the gel was dropped inside the bank consisting of Kapton tape. Gate voltages were applied via the ionic gel at 300 K to retain charged ionic molecular mobility in the ionic gel. After an electric double-layer was formed, the sample was cooled to 200 K to suppress the leakage current, and the HMR was measured. The accumulated charge density per area is typically $10^{14}$ cm$^{-2}$ [5]. The HMR measurements were performed with a Physical Property Measurement System (Quantum Design, USA) by applying a 30-μA dc electric current and changing the out-of-plane magnetic field *B* from −9 T to 9 T. Figure 1(b) shows the thickness dependence of the resistivity of the Pt films. In metallic thin films, electron scattering by surfaces and grain boundaries in the films non-negligibly contributes to carrier conduction. Thus, the thickness dependence of the resistivity can be described with the following equation [16],

$$\rho = \rho_{\text{bulk}} \left[ 1 - \left( \frac{1}{2} + \frac{3}{4} \frac{\lambda_{\text{mfp}}}{t} \right) \left( 1 - p e^{-\xi t / \lambda_{\text{mfp}}} \right) e^{-t/\lambda_{\text{mfp}}} \right]^{-1}, \quad (1)$$

where $\rho$, $\rho_{\text{bulk}}$, $\lambda_{\text{mfp}}$, $t$, $p$, and $\xi$ are the resistivity of the film, resistivity of the film in the bulk state, electron mean free path, thickness of the film, fraction of electrons specularly scattered at the surface, and grain–boundary penetration parameter, respectively. From the best fit with Eq. (1), we estimated $\rho_{\text{bulk}}$, $\lambda_{\text{mfp}}$, $p$, and $\xi$ to be 28 μΩcm, 14 nm, 0.89, and 0.42; comparable with those in a previous study [5].

Figure 2(a) shows the gate voltage dependence of the resistance *R*(0) of 1.7-nm-thick Pt under *B* = 0 T. Although Pt is a metallic material, we modulated the resistance by gating. The tendency of the resistance change is well explained by a conventional model of the ionic gate technique: upon application of a positive (negative) gate voltage, cations (anions) move to the top surface of the ultrathin Pt channel and form an electric double-layer (at the interface between the channel and ionic gel), which leads to electron carrier accumulation (suppression) and a decrease (an increase) of the resistance [17]. Note that the hysteresis behavior in the gate voltage dependence was originated from the polarization hysteresis of the ionic gel [17, 18]. Modulation of the resistance was ca. 19%, which signifies that the accumulated charge density is roughly estimated to be $10^{-21}$ cm$^{-3}$ because intrinsic carrier density of the ultrathin Pt is ca. $6 \times 10^{-21}$ cm$^{-3}$ [5]. The result can be rationalized by modulation of electronic states of ultrathin Pt by ionic-gate-induced carrier accumulation. Figure 2(b) presents the main result of this study: normalized HMR signals under gate voltages $V_G$ = 0, 1, and 2 V. We implemented the normalization by dividing the change of the resistance of the ultrathin Pt under application of an external magnetic field by the base resistance $R(0)$ at $B = 0$ T. A prominent finding is noticeable shifts of the amplitudes and changes in the shape of the HMR as a function of the gate voltage; i.e., HMR was larger under application of positive gate voltage in addition to a change of the spin lifetime (see also Supplementary Information regarding the reproducibility of the result in ultrathin Pt of various thicknesses).

To understand the physics that underlie the findings, we implemented the following analyses. The resistivity change of HMR is described by the equation as follows [9]:

$$\frac{R(B)-R(0)}{R(0)} = 2\theta_{SH}^2 \left\{ \left(\frac{\lambda}{t}\right) \tanh\left(\frac{t}{2}\right) - \text{Re}\left[\left(\frac{\Lambda}{t}\right) \tanh\left(\frac{t}{2\Lambda}\right)\right] \right\}, \quad (2)$$

where $R(B)$ is the resistance under $B$, $\theta_{SH}$ is the spin Hall angle, $\lambda = \sqrt{D\tau}$ is the spin diffusion length, ($D$ is the diffusion coefficient and $\tau$ is the spin lifetime), and $\Lambda = \left(\sqrt{\frac{1}{\lambda^2}+\frac{i}{\lambda_m^2}}\right)^{-1}$ with $\lambda_m = \sqrt{D/\Omega}$ ($\Omega = g\mu_B B/\hbar$ is the spin precession frequency, $g$ is the Lande factor, $\mu_B$ is the Bohr magneton, and $\hbar$ is the reduced Planck constant). Under a sufficiently strong applied magnetic field ($\Omega\tau \gg 1$), the second term of Eq. (2), $\text{Re}[(\Lambda/t)\tanh(t/2\Lambda)]$, can be regarded as zero. From this relationship, Eq. (2) can be described as follows [9,10] (see also Supplementary Information):

$$\frac{R(B)-R(0)}{R(0)} = \frac{R(\infty)-R(0)}{R(0)} \left\{ 1 - \left[\frac{1+\sqrt{1+(\Omega\tau)^2}}{2(1+(\Omega\tau)^2)}\right]^{1/2} \right\}, \quad (3)$$

The HMR signals under various gate voltages were nicely fitted by Eq. (3) [Fig. 2(b)]; which signifies that all of the results can be understood within the HMR scheme, whereas the shape and amplitudes of the HMR are gate-dependent.

The spin lifetime and spin Hall angle of the 1.7-nm-thick Pt at $V_G = 0$ V can be estimated by using Eq. (3) to be $\tau = 1.2$ ps, $\lambda = 1.9$ nm, and $\theta_{SH} = 0.023$; where $D$ is assumed to be $3.4\times10^{-6}$ m$^2$/s [9]. The estimated $\lambda$ and $\theta_{SH}$ are in accordance with a previous study ($\lambda = 1.5\pm0.5$ nm and $\theta_{SH} = 0.11\pm0.08$ [19]). Figures 3(a) and 3(b) show the entire dataset of the gate voltage dependence of $\tau$ and $\theta_{SH}$ of the 1.7-nm-thick Pt film, where $\tau$ decreased and $\theta_{SH}$ increased upon augmentation of the gate voltage. We observed a similar gate voltage dependence of HMR, $\tau$, and $\theta_{SH}$ in a thicker (3.3 nm) Pt film (see Supplementary Information). Whereas the gate voltage dependence of $\tau$ and $\theta_{SH}$ is reminiscent of enhancement of SOI under application of positive gate voltage, we discuss the underlying physics in the next paragraph.

Chemical reaction is a conceivable effect that could be brought by ionic gating to ultrathin Pt. Nevertheless, modulation of $R(0)$, $\tau$, and $\theta_{SH}$ was reversible in the experiments; which suggests that the contribution of chemical reactions was not dominant. Thus, modulation of the HMR stems from modulation of the electronic states at the top surface. As discussed by Velez *et al.* [9], HMR is a surface-sensitive effect because pure spin current (created close to the surface within the length scale of spin diffusion and arriving at the top/bottom surfaces) contributes to HMR. Furthermore, because we applied a strong electric field at the top surface by ionic gating, a nonnegligible Rashba field can be created at the interface between the ionic gel and ultrathin Pt. Suppression of the resistance by positive gating can predominantly occur at the top surface because of a screening effect of the gate voltages, which facilitates a selective increase of the charge current flow near the top surface (within < 1 nm from the top atomic layer). Consequently, the interplay of the enhanced Rashba field along the (-z) direction and charge current flow at the top surface can result in an enhanced HMR under application of positive gate voltage.

Of note is the difference between the findings of this study and the SOI suppression of ultrathin Pt by ionic gating that was corroborated by spin pumping from a YIG substrate beneath ultrathin Pt [5]. In ref. [5], the ISHE in ultrathin Pt was modulated by ionic gating, where modulation of the Fermi level due to charge accumulation attributed to the finding. Meanwhile, the modulation of the HMR is ascribed to the combination of suppression of the resistivity of the top surface adjacent to the ionic gate and creation of the Rashba field by the gating at the top surface. The more detailed physical picture is as follows: In that experimental scheme [5], pure spin current was constantly pumped from YIG under its ferromagnetic resonance to the bottom surface of the ultrathin Pt, and propagated in the Pt film along the direction perpendicular to the film plane. Most of the pumped pure spin current relaxed in the ultrathin Pt film before the spin current reached the top surface, because the spin diffusion length of the ultrathin Pt film was close to 1 nm [20]. However, in the present experimental scheme, pure spin current that contributes to HMR is created close to the top surface, whereas uniform charge current flow occurred in the ultrathin Pt film under zero-gating. Under positive gating, selective charge flow took place near the top surface by the screening effect and suppression of the resistivity under the positive gating. Although the total quantity of created spin current that contributes to HMR decreased by suppression of the bulk SOI of the Pt by gating, the Rashba SOI created by the gating can play a complementary role. The previous experimental scheme was not sensitive to such a complementary mechanism because most of the pumped spin current from the bottom surface did not reach the top surface.

In summary, we demonstrated gate-tunable HMR in ultrathin Pt films by using ionic gating, which facilitated efficient carrier accumulation in Pt and formation of an electric double-layer between the ionic gel and Pt. The Rashba field created by the gating plays the dominant role in modulating the HMR in addition to the suppression of the top surface resistivity in the ultrathin Pt by the gating. The findings of this study can facilitate investigations of gate-tunable spin-related effects in terms of the interplay of dense carrier accumulation (enables modulation of the electric states of an adjacent solid) and the interfacial Rashba effect (between the solid and ionic gate material by gating), which has not been shed light on in spintronics science. Furthermore, our findings can also facilitate fabrication of spin devices in which the spin functions are gate-tunable, given that modulation of physical effects by gating pioneered novel devices.

**Acknowledgments**

Part of this study was supported by a Grant-in-Aid for Scientific Research (S) No. 16H06330, "Semiconductor spincurrentronics,"

## Figures and Figure Captions

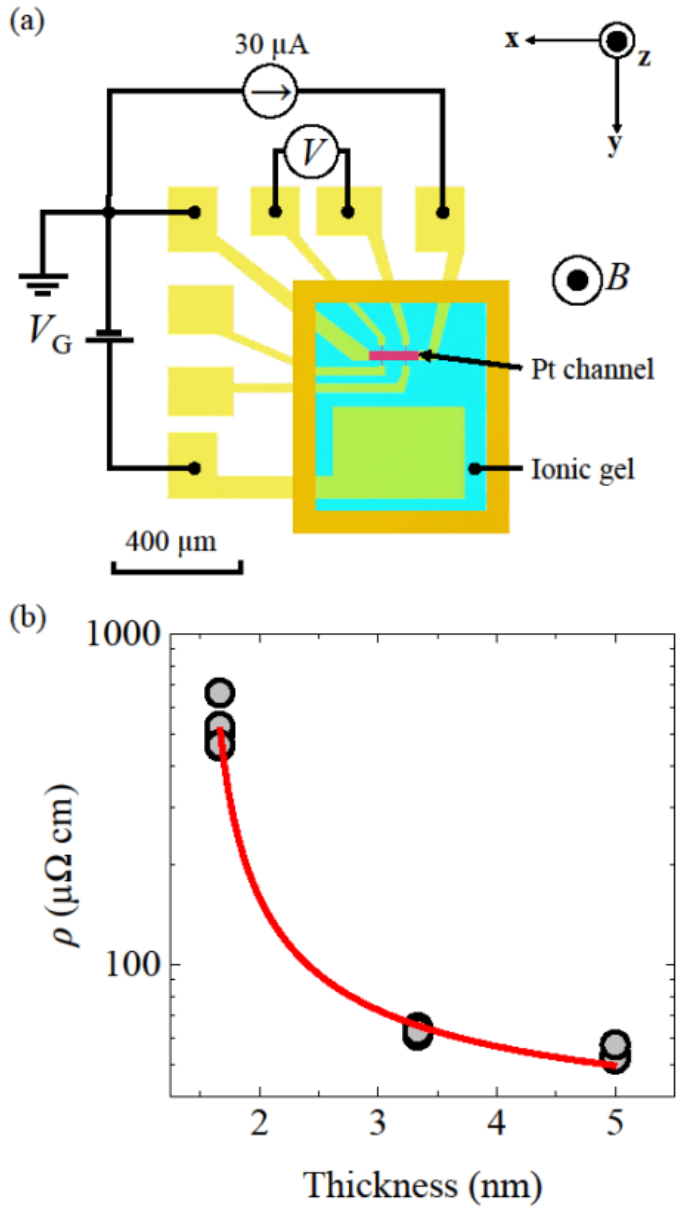

**Fig. 1.** (a) Measurement setup of gate-tunable Hanle magnetoresistance of ultrathin Pt grown on a GGG/YIG substrate. (b) Thickness dependence of the resistivity of ultrathin Pt films. Red solid line is the theoretical fitting line obtained with Eq. (1) in the main text.

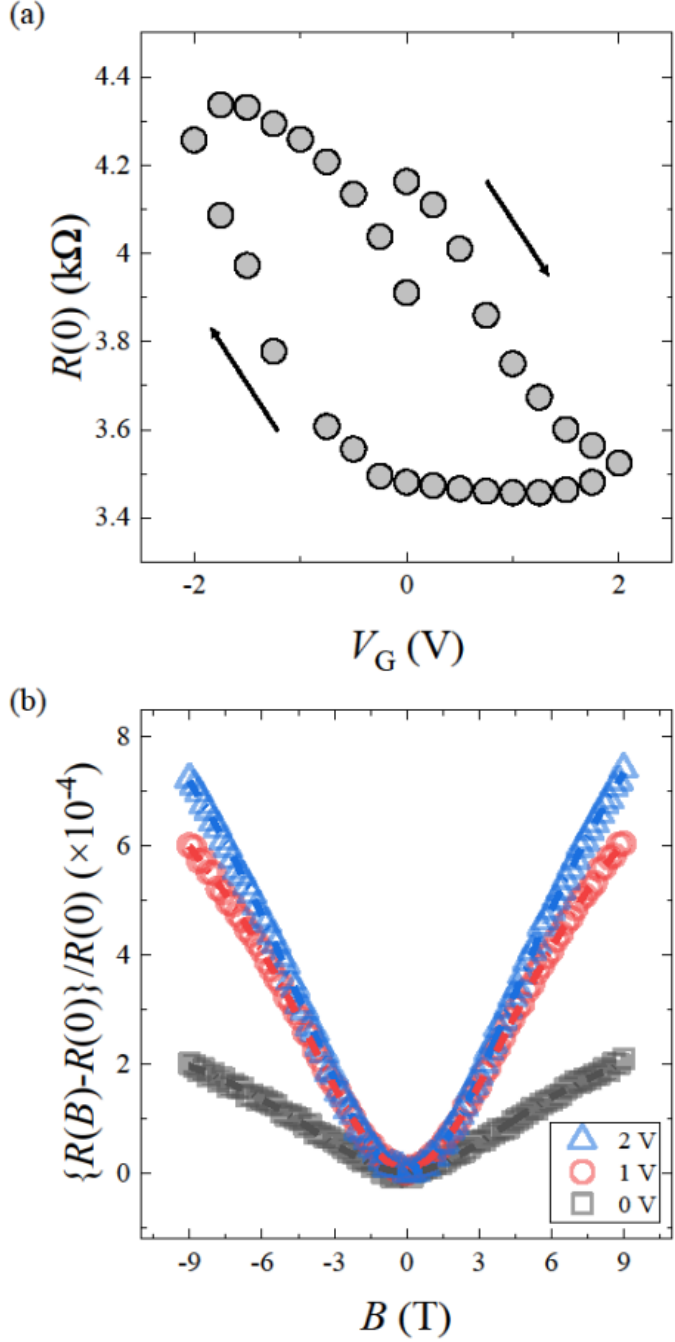

**Fig. 2.** (a) Gate voltage dependence of $R(0)$ (resistance under $B = 0$ T). Sweeping direction of the gate voltages is indicated with black arrows. (b) Normalized HMR signals under gate voltages of $V_G$ = 0 (black open squares), 1 (red open circles), and 2 (blue open triangles) V. Dashed black, red, and blue lines show the results of the theoretical fitting with Eq. (2) in the main text.

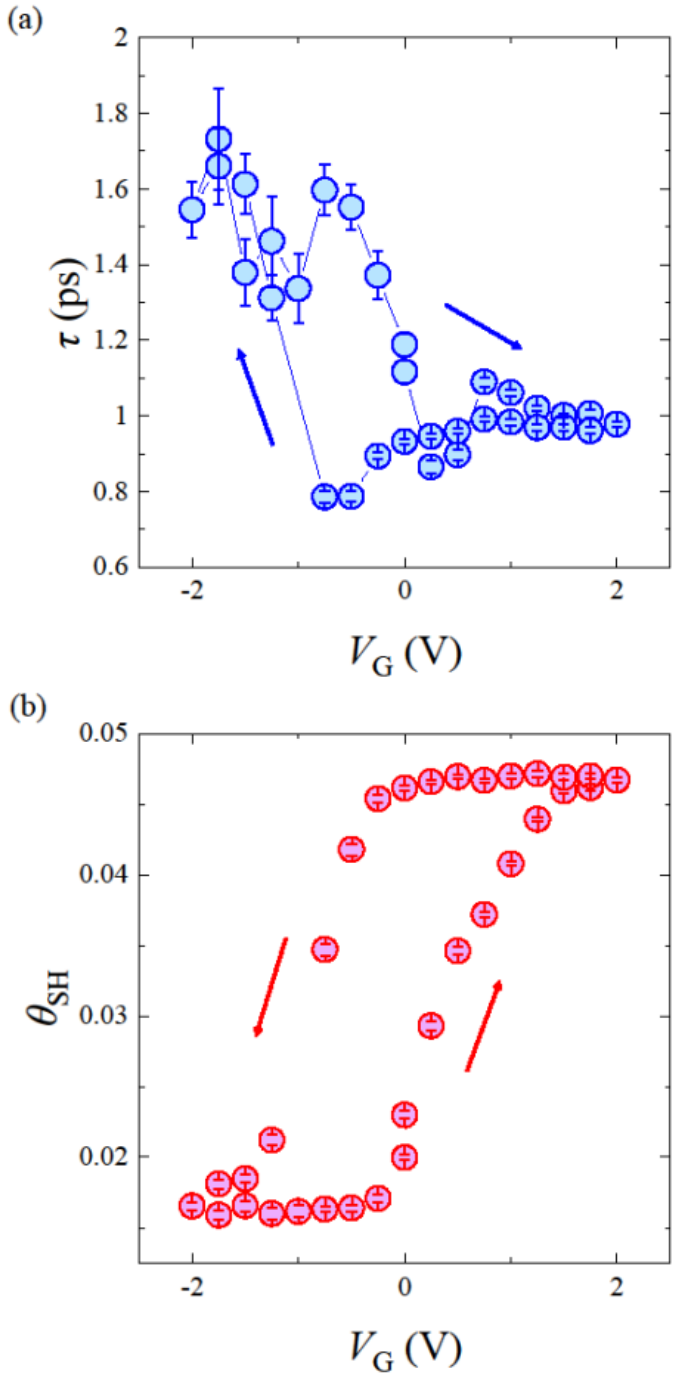

**Fig. 3.** Gate voltage dependence of the following: (a) spin lifetime $\tau$ and (b) spin Hall angle $\theta_{\mathrm{SH}}$. Estimated error bars in the fitting are shown. Sweeping directions of the gate voltages are also indicated with (a) blue and (b) red arrows.